\documentclass[pra,aps,floatfix,showpacs]{revtex4-1}
\usepackage{epsfig,graphicx,tabularx,epstopdf}

\begin{document}
\title{ Dynamics in asymmetric double-well   condensates}
\author{H. M. Cataldo and D. M. Jezek}
\affiliation{IFIBA-CONICET
\\ and \\
Departamento de F\'{\i}sica, FCEN-UBA
Pabell\'on 1, Ciudad Universitaria, 1428 Buenos Aires, Argentina}
%
%
\begin{abstract}
The dynamics of Bose-Einstein condensates in asymmetric double-wells is studied.
We construct a two-mode model and analyze the properties of the corresponding phase-space diagram,
showing in particular that the minimum of the phase-space portrait becomes shifted from the origin as a
consequence of the nonvanishing overlap between the ground and excited states
from which the localized states are derived.
We further incorporate effective interaction corrections in the set of two-mode model parameters.
Such a formalism is applied 
to a recent experimentally explored system, which is confined by  a toroidal trap 
with radial barriers forming an arbitrary angle between them.
 We confront the model results with Gross-Pitaevskii simulations for
various angle values finding a very good agreement.
We also analyze the accuracy of a previously employed simple model for moving barriers,
exploring a possible improvement that could cover a wider range of trap asymmetries.

\end{abstract}

\maketitle
\section{Introduction}

The two-mode (TM)  model has been studied and extensively
applied to symmetric double-well atomic Bose-Einstein condensates
in the last years 
\cite{smerzi97,ragh99,anan06,jia08,albiez05,mele11,abad11,doublewell,*xiong,*zhou,*cui,*gui}.
The dynamics of such a model  relies on assuming that the condensate order parameter can 
be described as a 
superposition of wave functions localized on each well with time-dependent coefficients.
The predicted Josephson and self-trapping regimes \cite{smerzi97,ragh99} have 
been  experimentally
observed by Albiez {\it et. al.}
\cite{albiez05}.
Also  ring-shaped condensates with two radial barriers forming a double well
 have been theoretically  studied in the case of  dipolar \cite{abad11}  
and contact \cite{je13}  interactions.

On the other hand, the situation is quite different in the case of asymmetric double wells,
since no similar theoretical development for such configurations has been reported so far.
This kind of systems  has become increasingly interesting due to
recent experiments \cite{boshier13,sato,jen14}, where  toroidal traps 
with two radial barriers moving symmetrically were investigated, which involves
a much richer type of dynamics. In fact, in addition to representing the first 
experimental realization of a SQUID analog with a Bose-Einstein condensate
\cite{boshier13,sato}, such experiments show a critical barrier velocity 
above which atoms become compressed on one side and
expanded on the other side,
in close analogy to the transition from dc to ac Josephson effects.
Such an ac regime presents a higher degree of complexity 
since excitations like solitons and vortices
may be shed into the rarefied portion of the condensate, giving rise at their eventual decay to
an additional resistive current \cite{jen14}. If we restrict ourselves to small displacements of the 
barriers from
the symmetric configuration, it has been shown in Ref. \cite{boshier13} that a straightforward
generalization of the symmetric TM model to such a dynamic configuration works well.
However, it is easy to understand that any first step to achieve an
 analogous model valid for more general barrier movements should
necessarily involve the study of an asymmetric TM model,
which constitutes the main goal of the present work. Thus, focusing on an arbitrary configuration
of fixed barriers, 
we begin in Sec. \ref{sec1}
by  analyzing the properties of the stationary states of an asymmetric pair of 
weakly coupled condensates, from which
both localized states are
derived. In fact,
the nonvanishing overlap between the Gross-Pitaevskii (GP) ground state and the 
excited stationary state turns out to  determine the position of the minimum in 
the phase-space diagram.
 We derive the full set of  TM model  parameters and 
further introduce in Sec. \ref{sec2} the  corrections in  the  interaction energy parameter 
using the proposal of Ref. \cite{je13} adapted for  an asymmetric system. 
In Sec. \ref{sec4.1}, we describe the system we use in our model applications and simulations
following the experimental settings of Ref. \cite{boshier13}.
The corresponding phase-space diagram is obtained in Sec. \ref{sec4.2}, where we
compare the TM results with  GP
 simulations, showing that our model, corrected by the modified effective  interaction parameters,
yields a much more accurate dynamics.
The case of moving barriers is considered in Sec. \ref{sec4.3}, where we
 derive  equations of motion similar to those 
employed in Ref. \cite{boshier13}, discuss the importance of terms 
disregarded in this approach and explore a possible improvement.
  Finally, in Sec. \ref{sec5} we summarize our work giving
some concluding remarks.

\section{Asymmetric two-mode model}\label{sec1}

The TM model  dynamics has been extensively studied in symmetric double-well 
potentials \cite{ragh99,anan06}. 
The commonly used ansatz
 for the wavefunction reads 
\begin{equation}
\psi_{TM}({\bf r},t)= b_{1}(t)\, \psi_{1}({\bf r}) + b_{2}(t)\, \psi_{2}({\bf r}),
\label{varan2m}
\end{equation}
where $ \psi_{1}({\bf r})$ and $ \psi_{2}({\bf r})$ are real, normalized to unity, 
localized wave functions at 
each well. The complex time-dependent coefficients are written as
$b_k(t)=\sqrt{N_k/N}\,e^{i\phi_k}$ ($k=1,2$), where $\phi_k$ and
$N_k$ represent the phase and particle number in the $k$ well, respectively,
and $N$ denotes the total number of particles.
We will use the same ansatz for our generic asymmetric configuration,
with localized states constructed from  the ground state   $\psi_{G}(\mathbf{r})$,
which we assume real and positive, 
   and from   the excited  stationary   state
 $\psi_E(\mathbf{r})$, which we assume real and negative  (positive) in the
`1' (`2') well.  
Both states are supposed to be normalized to 
one.
Then, the localized states read,
\begin{equation}
 \psi_{1}(\mathbf{r}) =  \frac{\psi_{G}(\mathbf{r})  -  \psi_E(\mathbf{r})}{\sqrt{2(1-\beta)}}
\end{equation}
\begin{equation}
 \psi_{2}(\mathbf{r}) =  \frac{\psi_{G}(\mathbf{r})  + \psi_E(\mathbf{r})}{\sqrt{2(1+ \beta)}}
\end{equation}
where $ \beta\equiv \langle \psi_{G} |  \psi_{E}\rangle=\int d^3{r}\, \psi_{G}(\mathbf{r})
\psi_{E}(\mathbf{r})$. Here 
it is important to remark that although the stationary states of the asymmetric case
present a nonvanishing overlap,
the corresponding localized states are indeed orthogonal by construction. 
Writing  the stationary states in terms of the localized ones we have,
\begin{eqnarray}
\psi_{G}&=&\sqrt{\frac{1-\beta}{2}}  \psi_{1}(\mathbf{r})  +  \sqrt{\frac{1+\beta}{2}} \psi_{2}(\mathbf{r}),\label{frpop1}\\
\psi_{E}&=&-\sqrt{\frac{1-\beta}{2}}  \psi_{1}(\mathbf{r})  +  \sqrt{\frac{1+\beta}{2}} \psi_{2}(\mathbf{r})\label{frpop2}.
\end{eqnarray}
Thus, we may see from the above
equations that both stationary states have identical
populations at each site.

As usual,  in order to obtain the TM dynamics, we introduce
 the order parameter
into the time-dependent GP equation,

\begin{equation}
i \hbar\frac{\partial\psi_{TM}({\bf r},t)}{\partial t}=
\left[-\frac{ \hbar^2 }{2 m}{\bf \nabla}^2  +
V_{\rm{trap}}({\bf r})+g\,N|\psi_{TM}({\bf r},t)|^2\right]\psi_{TM}({\bf r},t) \, .
\label{2t-dgp}
\end{equation}
Projecting it onto  $ \psi_{1}({\bf r})$ and $ \psi_{2}({\bf r})$,
and integrating  both  equations using
 the hopping and on-site energy parameters given in Eqs.  (\ref{epsR}) to  (\ref{ijota})
of the Appendix, one obtains
\begin{eqnarray}
 i \hbar\,\frac{db_{1}}{dt}& =&  \varepsilon_{1}  b_{1}  - K  b_{2}+U_{1} N|b_{1}|^2 b_{1}
-  F_{12} \, N  [2 {\rm Re}(b_{1}^*b_{2})b_{1} + b_{2}  | b_{1}|^2 ]
-  F_{21} N  b_{2}  | b_{2}|^2  \nonumber\\
&+& I N  [2 {\rm Re}(b_{1}^*b_{2})b_{2} + b_{1}  | b_{2}|^2 ],\\
 i \hbar\,\frac{db_{2}}{dt}& =&  \varepsilon_{2}  b_{2}  - K  b_{1}+U_{2} N|b_{2}|^2 b_{2}
-  F_{21} \, N  [2 {\rm Re}(b_{2}^*b_{1})b_{2} + b_{1}  | b_{2}|^2 ]
-  F_{12} N  b_{1}  | b_{1}|^2  \nonumber\\
&+& I N  [2 {\rm Re}(b_{2}^*b_{1})b_{1} + b_{2}  | b_{1}|^2 ].
\label{2mode0}
\end{eqnarray}
These equations  include all the terms introduced  for a symmetric system
in the improved two-mode model  \cite{anan06}.
In terms of  imbalance $ Z = (N_{2} - N_{1})/N $ and phase difference
 $ \phi= \phi_{1}- \phi_{2}$, we obtain the following equations of motion 
\begin{equation}
 \hbar  \dot{Z} =  -(J_a + Z J_b) \sqrt{1-Z^2}\sin\phi   +  I N  \,  (1 - Z^2) \sin (2 \phi) \, ,
\label{imbe}
\end{equation}
\begin{equation}
\hbar  \dot{\phi} = -A  + U_a  N Z   
+ J_ a \left[ \frac{Z}{\sqrt{1-Z^2}}\right]\cos\phi  + J_ b \left[ \frac{2 Z^2-1}{\sqrt{1-Z^2}}\right]\cos\phi
 -    I N \,  Z (2+  \cos (2 \phi)) \,,
\label{phasee}
\end{equation}
where 
\begin{equation}
A  =   \varepsilon_{1} - \varepsilon_{2} + \frac{1}{2}  N  ( U_{1}- U_{2})  \, ,
\label{A}
\end{equation}
\begin{equation}
J_a  = 2 K   + N  (  F_{21}  +  F_{12})  \, ,
\label{Ja}
\end{equation}
\begin{equation}
J_b  =   N  (  F_{21}  -  F_{12})  \, ,
\label{Jb}
\end{equation}
\begin{equation}
U_a  =  \frac {  U_{1}  +  U_{2}}{2}  \, ,
\label{Ua}
\end{equation}
with   the corresponding parameter definitions given in the Appendix.
The above equations of motion  can
 be obtained from the following Hamiltonian,
\begin{equation}
 H(Z,\phi)  =   -A Z  +  \frac{ U_a  N}{2}  Z^2 - (J_a + Z J_b) \sqrt{1-Z^2}\cos\phi  
 - IN Z^2 +  \frac{IN}{2}     (1 - Z^2) \cos (2 \phi) \, ,
\label{hamil}
\end{equation}
using the fact $Z$ and $ \phi $ are canonical conjugated coordinates i.e.,
$  \dot{Z} = -  {\partial H}/{\partial \phi }$ and
 $ \dot{\phi} =  {\partial H}/{\partial Z }.$ 

The phase-space portrait $ ( Z ,\phi) $ of such a Hamiltonian 
exhibits a minimum  $ ( Z_0 ,0) $, and  for strongly interacting systems,
a saddle $ ( Z_0 ,\pi) $ and two maxima $ (\pm Z_M ,\pi) $.
As can be easily deduced from Eqs. (\ref{frpop1})-(\ref{frpop2}),
 the overlap between the ground and excited states determines
 the stationary imbalance  $ Z_0=  \beta $. 
Thus,  we may see that minimum and  saddle 
 will be shifted in a $ Z= \beta $ value
from the corresponding locations on the $ \phi $ axis in the case of  the symmetric double well. 
Using the TM model parameters,  $Z_0$  may be   
approximated disregarding  almost negligible terms  by 
\begin{equation}
  Z_0  \simeq  \frac{ A }{ U_a N }    \, .
\end{equation}

The separatrix between closed, with a bounded phase (BP),  and open, 
with a running phase  (RP)  orbits  arises from the condition that the energy
corresponds to the saddle point, $ H ( Z, \phi) = H (Z_0, \pi)$.  Particularly,
for $\phi=0$ the separatrix yields a critical  imbalance $  Z_c$ given  approximately by
\begin{equation}
  Z_c  \simeq Z_0 \pm  \sqrt{\frac{ 4 (J_a+ Z_0 J_b) \sqrt{1-Z_{0}^2}}{ U_a N }}    \, ,
\label{critico}
\end{equation}
where the plus (minus) sign corresponds to the separatrix  above (below) the  BP orbits. 

\section{Effective interaction effects }\label{sec2}

The inclusion   of effective interaction
 effects in the TM model of a symmetric double-well system
 has shown to provide an accurate
correction to  the disagreements with GP simulations  \cite{cap13,je13}.
Such a correction  takes into account the density deformation 
of the GP mean-field term  when varying  the 
imbalance, which results in a net reduction of the interaction energy parameter.  
Within the Thomas-Fermi approximation,
the corresponding  reducing factor  can be analytically obtained \cite{cap13},
whereas a  numerical calculation using the ground state density 
has  shown to provide  accurate values  in more general cases \cite{je13}.
We note that a similar  correction  (about a 20 $\%$ reduction)  
has been introduced in the plasma oscillation frequency
of a tunable superfluid junction,  which was
shown to be crucial to accurately describing experimental results   \cite{LeBlanc11}.
  
 Here we follow the same
procedure of Ref.  \cite{cap13} and revise the term related to the interaction energy:
\begin{equation}
  U_{1}   N_{1}  -  U_{2}    N_{2}  \simeq    U^{\Delta N_{1}}_{1}   N_{1}  -  U^{\Delta N_{2}}_{2}    N_{2} \, ,
\label{renor1}
\end{equation}
where
\begin{equation}
 U^{\Delta N_{k} }_{k}    =   \int d^3r\,\,   \psi_{k}^2({\bf r}) \,   \rho_{k}^{\Delta N_{k}}({\bf r}) 
\,\,\,\,\,\,\,(k=1,2).
\label{urdelta}
\end{equation}
In  asymmetric double-well systems, in principle, the deformation of each localized density 
is not the same 
and thus  the effective interactions
$  U^{\Delta N_{1} }_{1}  $  and  $  U^{\Delta N_{2} }_{2}  $   \cite{cap13,je13}  should be 
treated separately.
The idea behind the method is 
that nonequilibrium states can be well aproximated by  localized on-site states
corresponding to the instantaneous population at each well, $N_k^0+\Delta N_k$,
where $N_k^0$ denotes the population of the $k$-well for $N$ particles.
Such localized states can be obtained from the stationary states  of systems
with a total number of particles different from $N$, whose
localized on-site densities (normalized to unity) are denoted by
$ \rho_{k}^{\Delta N_{k}}({\bf r}) $  in (\ref{urdelta}).
More details about this calculation will be given in Sec. \ref{sec4.1}.

We will assume that analogously to the symmetric case \cite{cap13,je13},
the following first-order approximation remains valid in any case
\begin{equation}
     U^{\Delta N_{k}}_{k}   =(1-2 \alpha_{k}  \frac{\Delta N_{k} }{N})  U_{k},
\label{renor2}
\end{equation}
where the parameter $\alpha_k$ may be numerically evaluated according to the procedure
of Ref. \cite{je13}.
Using  $  2 \frac{\Delta N_{1} }{N}= Z_0 -Z = -  2 \frac{\Delta N_{2} }{N}$
in   Eq. (\ref{renor2}) and replacing this result  in (\ref{renor1})   we obtain
\begin{eqnarray}
\frac{U^{\Delta N_{1}}_{1}   N_{1}  -  U^{\Delta N_{2}}    N_{2} }{N}
&=&  \frac{1}{2} [ (1- \alpha_{1} Z_0) U_{1} - (1+\alpha_{2} Z_0) U_{2})] \nonumber\\
&-&  \frac{Z}{2} [ (1- \alpha_{1} ) U_{1} +  (1- \alpha_{2}  ) U_{2})]
 -  \frac{Z (Z-Z_0)}{2}  (  \alpha_{1}  U_{1} - \alpha_{2}   U_{2}) .
\label{renor3}
\end{eqnarray}
Then, with the following definitions of effective on-site energy dependent  parameters,
\begin{equation}
\tilde{A}  =   \varepsilon_{1} - \varepsilon_{2} +    \frac{1}{2} N  [ (1- \alpha_{1} Z_0) U_{1} - (1+\alpha_{2} Z_0) U_{2}]    \, ,
\label{Amo}
\end{equation}
\begin{equation}
\tilde{U}_a  =    \frac{1}{2}   [ (1- \alpha_{1} ) U_{1} +  (1- \alpha_{2}  ) U_{2})]   \, ,
\label{Umo}
\end{equation}
\begin{equation}
\tilde{B }  =    \frac{1}{2}   (  \alpha_{1}  U_{1}  -  \alpha_{2}   U_{2} )   \, ,
\label{bmo}
\end{equation}
 and introducing the correction (\ref{renor3})  into the equation of motion (\ref{phasee}),
we obtain
\begin{eqnarray}
\hbar  \dot{\phi}  & =& -\tilde{A}   + \tilde{ U}_a  N Z    +   Z (Z-Z_0) N  \tilde{B}
+ J_ a \left[ \frac{Z}{\sqrt{1-Z^2}}\right]\cos\phi \nonumber\\  
&+& J_ b \left[ \frac{2 Z^2-1}{\sqrt{1-Z^2}}\right]\cos\phi
 -    I N \,  Z (2+  \cos (2 \phi)) \,,
\label{phaseemo}
\end{eqnarray}
which is consistent with the following `effective' Hamiltonian 
\begin{eqnarray}
\tilde{ H}(Z,\phi) & = & -\tilde{A} Z  +  \frac{\tilde{ U}_a  N}{2}  Z^2 - (J_a + Z J_b) \sqrt{1-Z^2}\cos\phi  
 - IN Z^2 \nonumber\\
&+&  \frac{IN}{2}   (1 - Z^2) \cos (2 \phi) 
-  (\frac{1}{2} Z_0 Z^2 - \frac{1}{3} Z^3 ) \tilde{B} N.
\label{rhamil}
\end{eqnarray}
The model represented by
the equations of motion (\ref{imbe})-(\ref{phaseemo}) and the Hamiltonian (\ref{rhamil}),
will be called as the effective two-mode (ETM) model in what follows.
With respect to the new phase-space portrait derived from this Hamiltonian,
it can be easily verified that the position of the minimum  remains located at $Z_0$,
which may be also approximated by
\begin{equation}
  {Z}_0  \simeq  \frac{ \tilde{A} }{ \tilde{U}_a N }  ,
\label{min}
\end{equation}
whereas the shape of the orbits may differ  from  that obtained with
the bare parameters, as will be shown in the following sections.
We also note that $\tilde{U}_a$ and $\tilde{A}$ become reduced with respect to 
$U_a$ and $A$, as seen from
 Eqs. (\ref{Amo})  and (\ref{Umo}), provided the parameters $\alpha_k$ are positive.
On the other hand, we remark that
the parameter $\tilde{B}$ arises from the combined effects of interaction and asymmetry.

\section{Numerical results}

\subsection{The system}\label{sec4.1}
We describe in what follows the system utilized in our simulations and model applications.
All the trapping parameters and condensate details have been chosen to reproduce the experimental setting of Ref. \cite{boshier13}.
The trapping potential can be written as the sum of a part that depends only
on $x$ and $y$ and a part that is harmonic in the tightly bound direction $z$:
\begin{equation}
V_{\text{trap}}(x,y,z)=V(x,y)+\lambda^2z^2
\end{equation}
being
\begin{equation}
V(x,y)=V_{\text{T}}(r)+V_{\text{B }}(x,y).
\end{equation}
The above potential consists of a superposition of
a toroidal term $V_{\text{T}}(r)$  ($r^2=x^2+y^2$)
and the radial barrier term   $V_{\text{B }}(x,y)$.
The toroidal potential was modeled through the following Laguerre-Gauss optical potential
\cite{lag}
\begin{equation}
V_{\text{T}}(r)=-V_0\left(\frac{r^2}{r_0^2}\right)\,\exp\left(1-\frac{r^2}{r_0^2}\right),
\label{toro}
\end{equation}
where $V_0$ corresponds to the depth of the potential and $r_0$ the radial position
of its minimum.
We have used scaled units referenced to a chosen unit of length denoted by $L_0$
(in our case $L_0$=1 $\mu$m). Energy and time units were defined in terms of $L_0$:
\begin{equation}
E_0=\frac{\hbar^2}{mL_0^2},\,\,\,T_0=\hbar/E_0,
\end{equation}
where $m$ denotes the mass of a condensate atom. 
For the present case of $^{87}$Rb atoms 
we have $E_0/k_B=5.5298$ nK and $T_0=1.3813$ ms.

\begin{figure}
\includegraphics{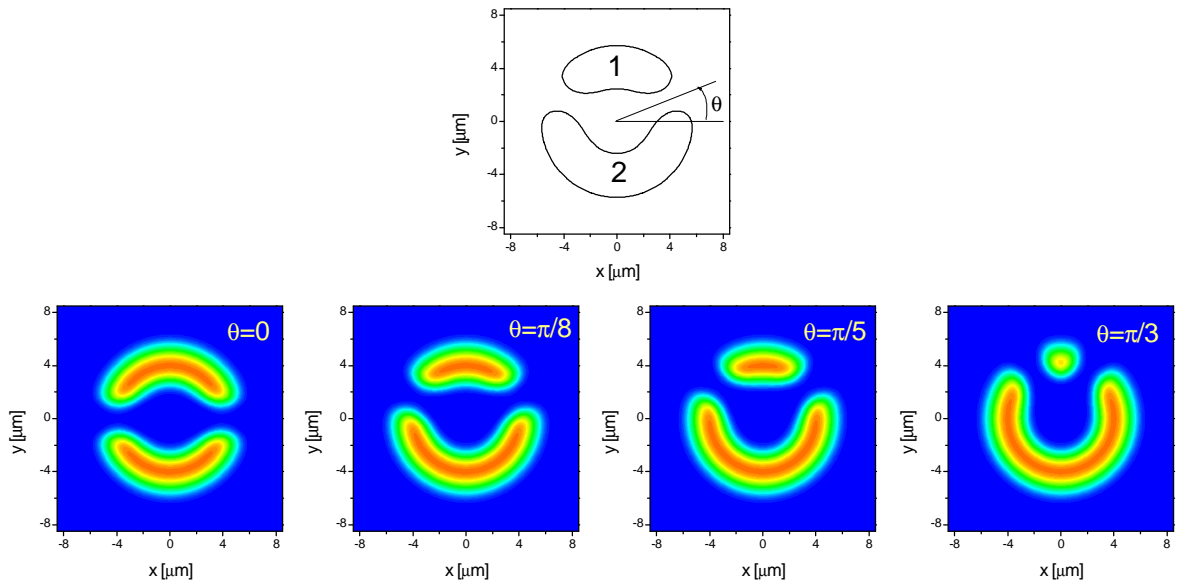}
\caption{(Color online) Particle density isocontours for the ground state at different positions
of the barriers for $N$=3000.}
\label{figu1}
\end{figure}

The barrier was modeled as
\begin{equation}
V_{\text{B}}(x,y) =
V_b \,\, \sum_{k=1}^{2} \exp \left\{ -\frac{ [ y\cos\theta_k 
- x\sin\theta_k ]^2}  
{ \lambda_b^2}\right\} \Theta[y \sin\theta_k + x\cos\theta_k  ],
\label{barre}
\end{equation}
where $ \Theta $ denotes the Heaviside function with $\theta_1= \theta $ and 
$ \theta_2= \pi-\theta$. The parameter $\theta$ will be assumed as time dependent
in the case of moving barriers. 
We have utilized, according to Ref. \cite{boshier13}, the following trap parameters:
  $V_0$=70 nK, $r_0$=4 $\mu$m,
$ V_b  =  41.07 $ nK, and $ \lambda_b = 1\, \mu$m. These barrier parameters
yield a full-width at half-maximum of the barrier nearly
below 2 $\mu$m, which is in agreement with the experimental data leading to a tunnel junction. 
We have assumed a high $\lambda= 8$ value yielding
a quasi-bidimensional condensate and allowing a simplified
numerical treatment \cite{castin}.
So, stationary states are written as the product
of a two-dimensional (2D) wave function $\varphi(x,y)$
and a Gaussian wave function along the $z$ coordinate,
$\sqrt{\frac{\lambda^{1/2}}{\pi^{1/2}}}\,\,e^{-\frac{\lambda z^2}{2}}$.
Thus, assuming barriers remaining at rest,
the GP equation for the former reads  \cite{castin}
\begin{equation}
-\frac{1}{2}\left(\frac{\partial^2\varphi}{\partial x^2}+\frac{\partial^2\varphi}{\partial y^2}\right)+
V(x,y)\,\varphi+gN\sqrt{\frac{\lambda}{2\pi}}\,|\varphi|^2\varphi=\mu \varphi
\label{gp}
\end{equation}
with
\begin{equation}
g=\frac{4\pi\hbar^2a/m}{E_0L_0^3}=4\pi a/L_0,
\end{equation}
where $a= 98.98\, a_0 $ denotes the  
$s$-wave scattering length of $^{87}$Rb, $a_0 $ being the Bohr radius.
In   Fig.~\ref{figu1} we depict the 2D particle  density $|\varphi(x,y)|^2$ for the ground state
 at different positions
of the barriers. According to the notation of previous sections
we will call the top and bottom wells of Fig. 1 as `1' and `2', respectively.
In the following we will restrict our calculations to a system with $N=3000$. 
As regards the excited state, we have obtained its wavefunction by evolving in imaginary time an
initial wavefunction identical to that of the ground state in site `2', whereas we introduced a change
of sign in the site `1', i.e.,  a wavefunction positive in site `2' and negative in site `1', a feature that 
turns out to persist until final convergence to
the stationary excited state. This procedure works well for all configurations below 
$\theta= 0.394 \pi$, while
for more asymmetric systems, the energy gap between the excited and the ground 
states becomes so low that
 the imaginary-time evolution leads to a `decay' to 
the ground state. Here it is interesting to notice that,
generalizing the TM result of the symmetric case \cite{LeBlanc11},
 such a gap
reads $(J_a + Z _0J_b) \sqrt{1-Z_0^2}>0$, which vanishes for $Z_0\rightarrow 1$, as expected.

\begin{figure}
\includegraphics{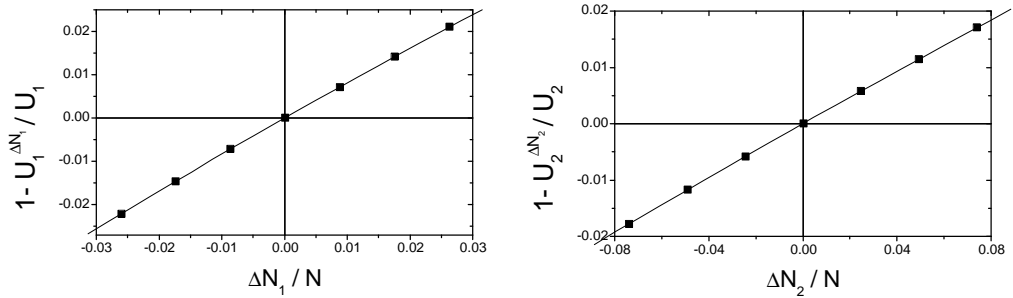}
\caption{The functions $ 1-  U^{\Delta N_k}_k /U_k$ for top well (left panel)  and
bottom well (right panel), versus each corresponding 
particle number difference for $\theta=\pi/5$. 
The square dots represent calculated values, while the solid lines
correspond to linear fits of such values.}
\label{figu3}
\end{figure}

To illustrate the method we used to calculate
the coefficients $\alpha_k$ in Eq.~(\ref{renor2}), we depict in Fig.~\ref{figu3}
the quantities $ 1-  U^{\Delta N_{k}}_{k} /U_{k}$
as functions of $  \frac{\Delta N_{k} }{N} = (N_{k} -N_{k}^0)/N $  for $\theta = \pi/5$,
where $N_{k}^0$ is obtained from
the projection of the ground state  
onto the  $k$-localized state for $N=3000$, whereas
$N_{k}$ is calculated analogously, but with a different
total number of particles. 
Thus,  according to Eq.~(\ref{renor2}), we have extracted the values of
  $\alpha_{k}$ from the slope of the lines. 
We note that $ \alpha_{1}$  turns out to be larger than $ \alpha_{2}$ because 
the smaller condensate should present the larger deformation
for an identical change in the particle number.

\subsection{Phase-space portrait and dynamics for static barriers }\label{sec4.2}

\begin{figure}
\includegraphics{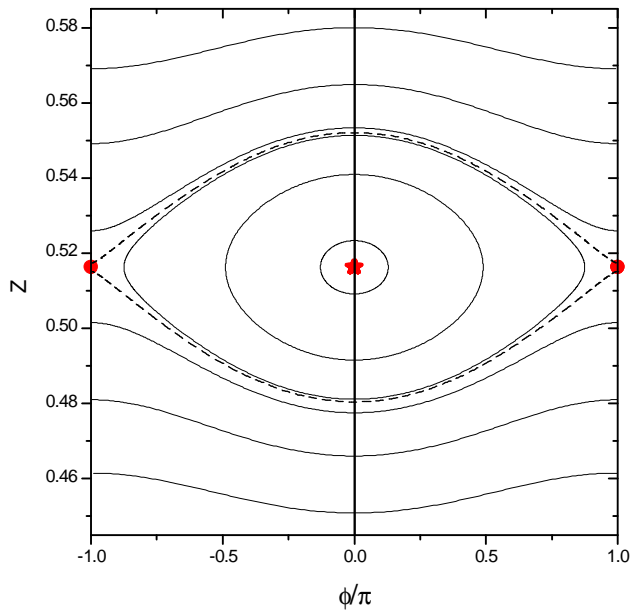}
\caption{(Color online) Phase-space portrait $Z$ versus $\phi$ for $\theta= \pi/5 $ 
arising from
Hamiltonian (\ref{hamil}). Each orbit is represented by a solid line, 
except for the separatrix between
 closed and open orbits, which is represented by the dashed line.
The minimum has been indicated by a red star and the saddle point by  red circles.}
\label{figu2}
\end{figure}

  We first note  that for $ 0 \leq \theta \leq \pi /2 $, 
the overlap  between the ground and excited states verifies
 $ 0 \leq  \beta \leq 1$, and the same occurs for 
the position of minimum and saddle since $Z_0=\beta$. 
In Fig. \ref{figu2} we depict the phase-space diagram of 
Hamiltonian (\ref{hamil}) for $\theta= \pi/5 $. 
In this case we have $Z_0=0.5163$,
which corresponds to the $Z$ coordinate of minimum and saddle in such a figure.
The separatrix between BP  and RP orbits has been numerically obtained 
and has been denoted by dashed lines in Fig. \ref{figu2}. 
We remark that the values $ Z_c= Z_0 \pm 0.0358$ derived from Eq. (\ref{critico})
 are in well accordance  with
the intersections of the dashed lines and the vertical axis. 

\begin{figure}
\includegraphics{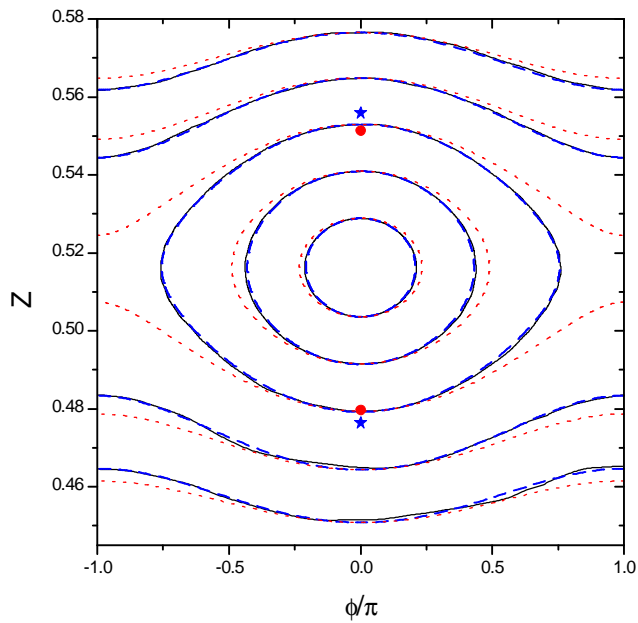}
\caption{(Color online) Imbalance $Z$ versus phase difference $\phi$  for $\theta= \pi /5$.
The GP simulation results are
represented by black solid lines and the ETM model
results by blue dashed lines. Orbits arising from the bare TM model 
are also depicted using  red dotted lines.
The separatrix points ($Z=Z_c,\phi=0$) are indicated by dots:
red circles and blue stars correspond to TM and ETM models, respectively.  
}
\label{figu4}
\end{figure}

In Fig. \ref{figu4} we depict the phase-space portrait arising from GP simulations and from ETM
and TM models.
We notice that the minimum of Hamiltonian (\ref{rhamil})
 remains located at $Z_0$ (cf. Eq. (\ref{min})),
whereas the shape of the orbits  differ from
 that observed in Fig. \ref{figu2}. Particularly, 
the separatrix between closed and open orbits covers a wider range of $Z$ values,
as shown in Fig. \ref{figu4}
through the locations of the critical imbalance $Z_c$ arising from the TM model (cf. Eq. 
(\ref{critico})) indicated by red dots, and those arising from the ETM model (blue stars).
Here it is worth noticing also that the value of $Z_c$ obtained from GP simulations coincides
with the corresponding ETM result.
Therefore, we may conclude that the matching between ETM model and GP simulation
results turns out to be 
much better than that of the plain TM model. 
\begin{figure}
\includegraphics{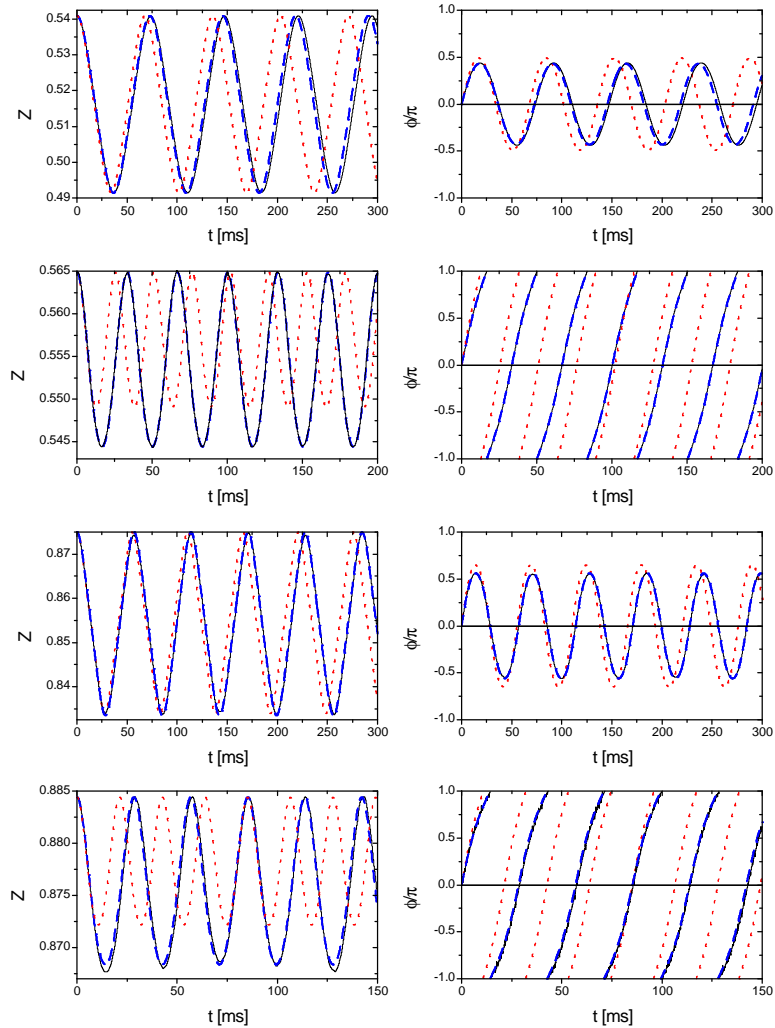}
\caption{(Color online) Time evolution of the imbalance $Z$ (left panels)
and the phase difference $\phi$ (right panels) for $\theta=\pi/5$ 
(upper four panels) and $\theta=\pi/3$  (lower four panels). The GP simulation results are
represented by black solid lines, the bare TM model results by red dotted lines, and the ETM model
results by blue dashed lines.  }
\label{figu5}
\end{figure}
 
We have depicted  in Fig. \ref{figu5} the time evolution of imbalance and phase difference
for $\theta =\pi / 5$ and $\theta =\pi / 3$
arising from GP simulations, together with the corresponding TM and ETM results. 
It is remarkable that the excellent agreement
between GP and ETM time evolutions persists even for most asymmetric configurations.

\subsection{ Moving barriers }\label{sec4.3}

The experimental results of Ref. \cite{boshier13} were well reproduced
from a simple model \cite{giova00},
 that adapted the TM equations of motion of a symmetrical configuration to
the case of moving barriers by simply taking into account the effect of a $\theta$-dependent equilibrium
imbalance $Z_0(\theta(t))$. Such a simple model for moving barriers (SMMB),
can be easily derived from our equations of motion
(\ref{imbe})-(\ref{phaseemo}) by approximating all the model
parameters by their values of the symmetric configuration, neglecting terms proportional to
the small parameter $I$, and replacing $\tilde{A}\rightarrow Z_0(\theta(t))\tilde{U}_a N $
 according to (\ref{min}). Thus we obtain,
\begin{equation}
 \hbar  \dot{Z} =  -J   \sqrt{1-Z^2}\sin\phi    \, ,
\label{simbe}
\end{equation}
\begin{equation}
\hbar  \dot{\phi} =  \tilde{ U}  N [ Z - Z_0(\theta(t))]
+ J   \frac{Z}{\sqrt{1-Z^2}}  \cos\phi   \, ,
\label{sphasee}
\end{equation}
where $ J= J_a $ and 
$ \tilde{ U}= \tilde{ U}_a $ respectively denote hopping and on-site energy
 parameters given by the corresponding values
of the symmetric case ($\theta=0$).  The barrier movement in these equations is 
represented by $Z_0(\theta(t))$, and of course only small departures from
the symmetric configuration should be expected to be well reproduced. 
Particularly, the experiments
in \cite{boshier13} were restricted to $\theta<\pi/8$, and
in this paper we will explore the dynamics for a wider range of
asymmetric final configurations.

By numerically  analyzing   $Z_0$  as a function of  $\theta$,  we 
have observed a linear behavior $Z_0=\alpha\,\theta$ ($ 2 \pi \alpha = 5.1585$),
except for values reaching $\theta\simeq 0.4\pi$
where both barriers begin to overlap ($Z_0\rightarrow 1$).
We note that the approximation
$ 2 \pi \alpha = 4$ used in Ref.  \cite{boshier13}, which amounts to assuming a linear behavior
of $Z_0(\theta)$ up to $\theta=\pi/2$ ($Z_0(\theta=\pi/2)= 1$),
 corresponds to the limit of a negligible
 barrier width and also
neglecting any healing length arising from the presence of barriers.

For simplicity, we will assume in this paper
barriers moving with a constant angular frequency $f_b$, thus  $ \theta(t)= 2 \pi f_b\,t$ and
we may approximate
\begin{equation}
 Z_0(\theta(t))  =  \alpha 2 \pi f_b\, t .
\label{z0}
\end{equation}

\begin{figure}
\includegraphics{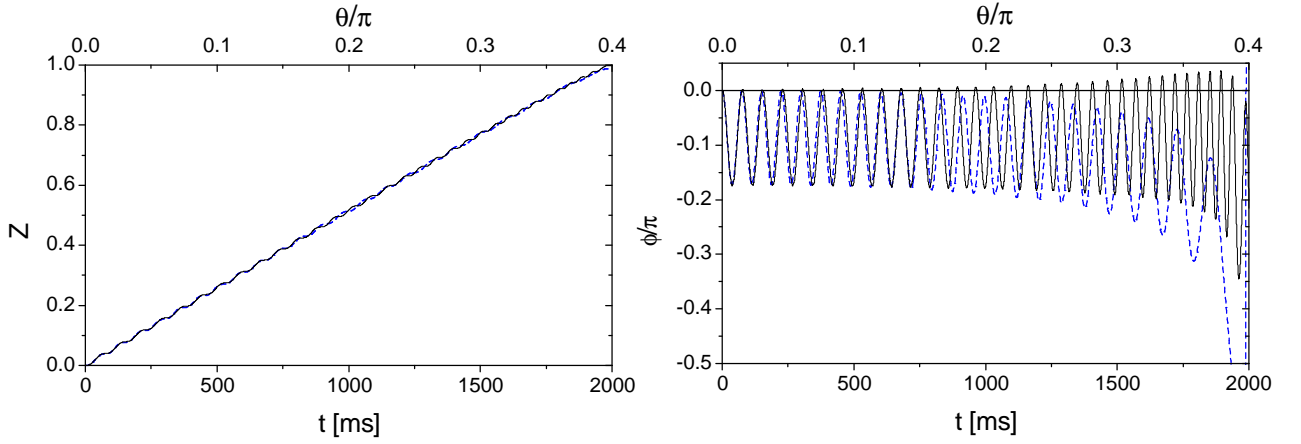}
\caption{(Color online) Time and $\theta$ dependence of the imbalance $ Z$ (left panel) and 
the phase difference $\phi$ (right panel) for
the barrier angular frequency $f_b=0.1$ Hz.
The GP simulation results are represented by black solid lines and the SMMB
results by blue dashed lines.}
\label{figu6}
\end{figure}

\begin{figure}
\includegraphics{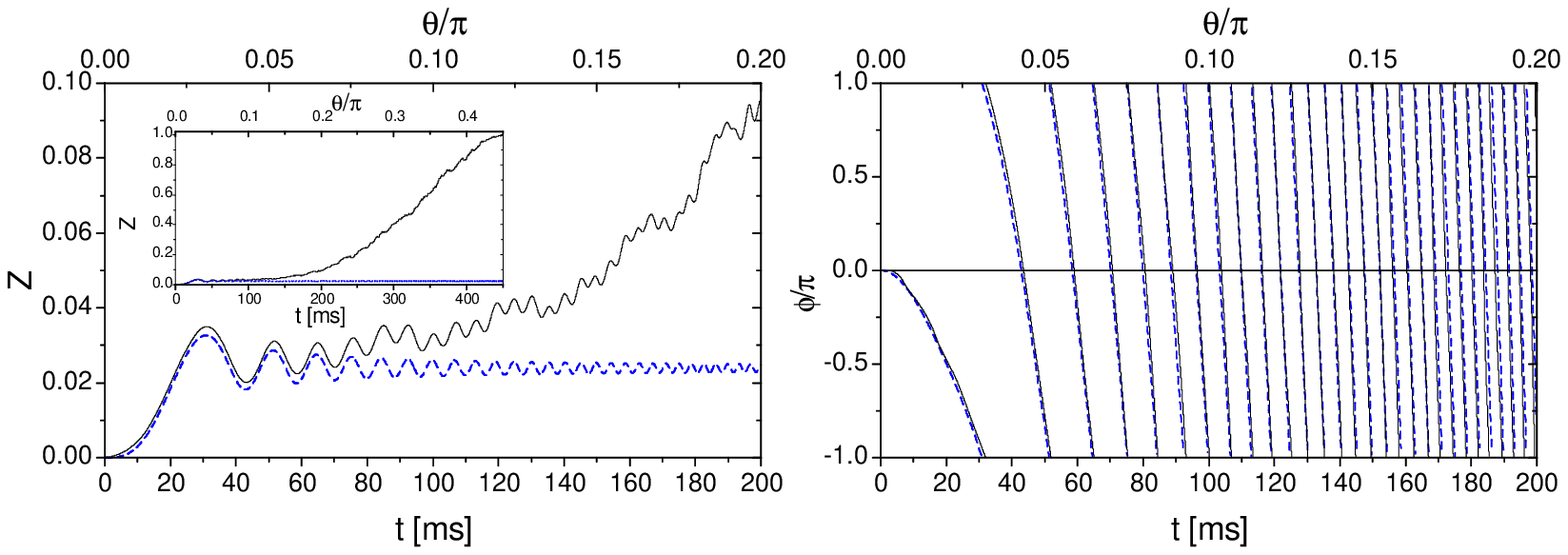}
\caption{(Color online) Same as Fig. \ref{figu6} for
the barrier angular frequency $f_b=0.5$ Hz. The inset within the left panel corresponds to a
more extended evolution of the imbalance.
}
\label{figu7}
\end{figure}

In Figs. \ref{figu6}  and \ref{figu7} we depict the time evolution of imbalance
$ Z$ and phase difference $\phi$ for barriers moving with $f_b=0.1$ Hz and $0.5$ Hz, 
respectively. The solid lines correspond to GP simulation results, while the dashed ones 
correspond to the SMMB results.
 There we may observe typical behaviors of the so-called dc-
and ac-Josephson regimes for barrier frequencies 0.1 and 0.5 Hz,
respectively  \cite{giova00,boshier13}. In fact,
the dc-Josephson regime,  which occurs below certain
critical barrier frequency,  is characterized by small oscillations around 
$Z_0(t)$ and a  bounded phase difference,  and for this reason we shall call it as
the dynamical bounded-phase (DBP) regime.
On the other hand, for a large
enough barrier velocity  a compression dynamics occurs \cite{boshier13,jen14},
which defines the ac-Josephson regime. Such a
regime exhibits an unbounded phase and so we shall call it in what follows 
as the dynamical running-phase (DRP)  regime.

It is instructive to analyze the short-time behavior of both regimes  using  the SMMB equations
(\ref{simbe})-(\ref{sphasee}).
For small barrier angular frequencies
within the DBP regime,  an analytically  tractable dynamics arises  by linearizing the 
SMMB for   $Z\ll1$ and  $\phi\ll1$. Thus
we obtain the approximate solutions,
\begin{equation}
Z  =  2 \pi \alpha f_b [t - \frac{1}{\omega_p} \sin( \omega_p t)]  \, ,
\label{soimbe}
\end{equation}
\begin{equation}
\phi =  -\frac{   \hbar 2 \pi \alpha f_b  }{J} [1 -  \cos ( \omega_p t) ]   \,,
\label{sphase1}
\end{equation}
with  $ \omega_p  = \sqrt{  \frac{\tilde{ U}  N  J }{\hbar^2} } $,
that qualitatively resembles the GP dynamics of Fig. \ref{figu6},  mainly the fact that $Z$ oscillates 
around $Z_0(t)$ and  the bounded phase remains confined to negative values.  
For longer times, increasing differences between the GP simulation results and those of the SMMB
are observed for $\theta> 0.15$ $\pi$ 
in the phase evolution at the right panel of Fig. \ref{figu6}.

\begin{figure}
\includegraphics{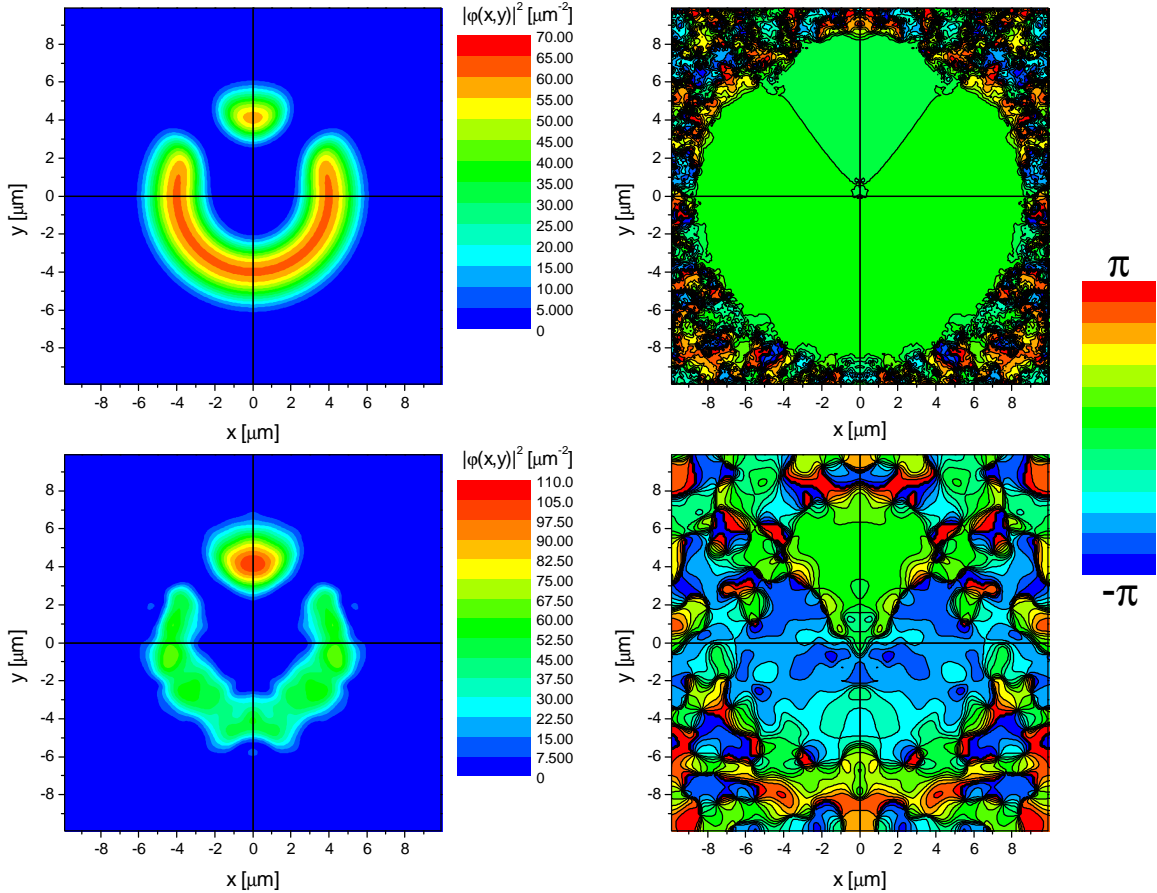}
\caption{(Color online) Snapshots of the particle density (left) and phase distribution (right).
The top
panels correspond to $f_b=0.1$ Hz and $t=1500$ ms, while the bottom panels correspond
to $f_b=0.5$ Hz and $t=300$ ms. In both cases the barriers are located at $\theta=0.3 \pi$. }
\label{figu11}
\end{figure}

On the other hand, in the  DRP regime, we may approximate
for large barrier angular frequencies
\begin{equation}
\hbar  \dot{\phi}  \simeq  -  \tilde{ U}  N  Z_0(t) =     -  \tilde{ U}  N   2 \pi \alpha f_b t \, ,
\label{sphaseer}
\end{equation}
which yields an unbounded  monotonically  decreasing phase. This qualitatively reproduces the
behavior of the phase difference shown in the right panel of Fig.~\ref{figu7}, 
particularly the increasing negative slope that is observed along the evolution. 
As regards the imbalance shown in the left panel,
we may see that the GP behavior is only reproduced by the SMMB at very short times ($t<60$ ms).
For longer times the SMMB completely fails to describe the GP dynamics. 
In particular, from the inset in this figure,
it may be seen that the model asymptotically  oscillates around  $ Z \simeq 0.02$, 
whereas the GP simulation shows an almost quadratic behavior,  approaching $ Z = 1 $. 
Note that any realistic dynamics
should tend to  $Z=1$   where the barriers are superposed.
The above failure of the SMMB in the DRP regime is easily understood from a simple inspection
of the density and phase snapshots shown in Fig. \ref{figu11}. In fact, it is clear that
any variant of the TM model, like that yielding the
SMMB, is supposed to rely on assuming that the shape of the density should at any time
 resemble that of a stationary state (Fig.~\ref{figu1}), and the phase should remain 
almost homogeneous at each well. This is indeed the case for the snapshots of $ f_b=0.1$ Hz in Fig. \ref{figu11}, which
are consistent with the quite acceptable results for the SMMB in Fig. \ref{figu6}, and also
suggest that
an `improved SMMB' covering a wider range of  asymmetric
configurations could be eventually devised for the DBP regime. 
On the other hand, the snapshots of $ f_b=0.5$ Hz show a very different situation, with important
deformations in the density, 
as compared to that of the ground state, and with clear inhomogeneities
in the phase at the bottom well, which reflects the formation of excitations like vortices
in the rarefied portion  of the condensate, as also discussed in Ref. \cite{jen14}.
None of these features could be taken into account in
any simple model like the SMMB.

\begin{figure}
\includegraphics{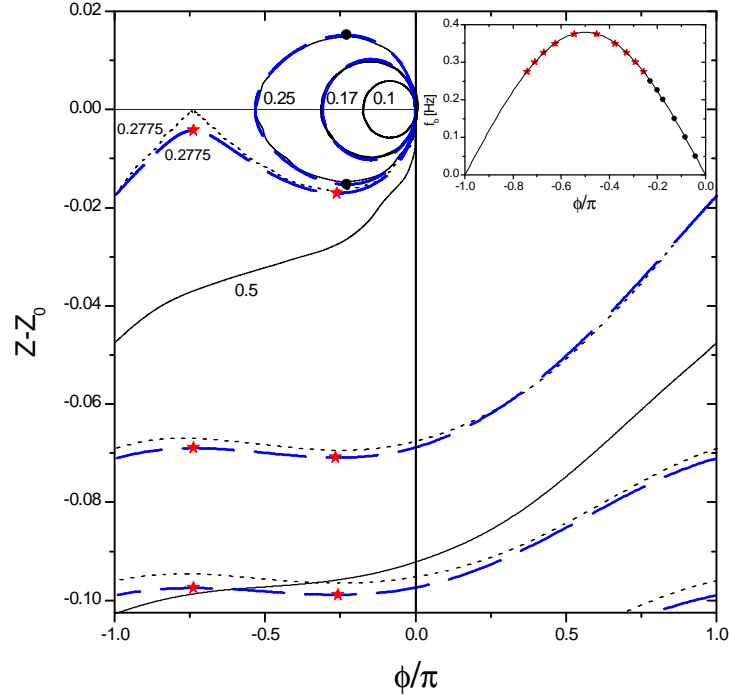}
\caption{(Color online) $ Z - Z_0$ versus $\phi$ for several barrier velocities. 
The GP simulation results (black solid lines) may be compared to 
the SMMB results (blue dashed lines) for the barrier frequencies
0.17, 0.25 and 0.2775 Hz. The GP critical barrier frequency
$f^{(2)}_b\simeq 0.2775$ Hz is represented by black dotted lines. The SMMB $Z-Z_0$
turning points for 0.25 Hz (0.2775 Hz) are represented by black circles (red stars).
Inset: the phase of the first SMMB $Z-Z_0$
turning points versus the barrier frequency for  DBP (DRP) evolutions
is represented by black circles (red stars), while the estimate (\ref{turnp}) is 
depicted by a solid line.}
\label{figu8}
\end{figure}

 In Fig.~\ref{figu8}   we depict $ Z(t) - Z_0(t)$ as a function of $\phi(t)$ obtained from
GP simulations and from the SMMB for several frequencies. 
It may be
seen that  the DBP  oscillations are confined within an oval-like region, 
whereas the DRP evolutions are localized
on the negative $ Z-Z_0 $ half-plane and can acquire  any phase value.
As regards the DBP oscillations, it is important to remark that they are not
closed orbits, since each new loop does not exactly reproduce the previous one.
Moreover, after a number of DBP loops, a given orbit (e.g., $f_b=0.25$ Hz) 
may `decay' to the DRP regime.
So, for the sake of clarity, we have only plotted  in Fig.~\ref{figu8}
that part of the trajectory corresponding to
the first period of each DBP evolution.
In addition, we point out that the asymmetry parameter $\theta$ in Fig.~\ref{figu8}  
does not exceed in any case 
the experimental limit of $\pi/8$.

It is interesting to analyze the turning points of the above trajectories, where
the time derivatives of $ Z-Z_0 $ or $ \phi$ vanish.  
The $ \phi$ turning points are located, as seen in Fig.~\ref{figu8},
 approximately at the origin and on the negative $ \phi$ axis for the DBP
loops, while they are absent in the DRP regime, as expected. On the other hand,
the $ Z-Z_0 $ turning points change from being a couple of almost
vertically aligned points for the DBP loops
(e.g. the black circles for $f_b=0.25$ Hz in   Fig.~\ref{figu8}), to becoming pairs of 
a maximum at left and a minimum at right,
located almost
symmetrically with respect to the vertical line $\phi=-\pi/2$, as indicated by the red stars
for $f_b=0.2775$ Hz. However, there are no such turning points above certain
frequency, as observed for $f_b=0.5$ Hz in   Fig.~\ref{figu8}. This may be deduced from the
following estimate of the phase $\phi$ of the $Z-Z_0$
turning points
\begin{equation}
f_b\simeq -\frac{J\sin\phi}{\hbar2\pi\alpha},
\label{turnp}
\end{equation}
which stems from Eqs. (\ref{simbe}) and (\ref{z0}) using the approximation $Z^2\ll 1$.
First it is convenient to compare the above prediction with the phase
of the first SMMB turning points for several barrier frequencies in the inset of Fig.~\ref{figu8}.
We may see that there is an excellent
agreement between the above formula and such points, while we have found
that this approximation remains quite acceptable for subsequent periods within 
$\theta < \pi/8$, as observed for the red stars in the main plot of Fig.~\ref{figu8}.
Now, one may define two critical frequencies $f^{(1)}_b$ and $f^{(2)}_b$
from the inset of Fig.~\ref{figu8}.
In fact, above the maximum of the sinusoid $f^{(1)}_b\simeq J/(\hbar2\pi\alpha)\simeq 0.38$ Hz,
there are no more turning points, which corresponds to the barrier frequency leading to the
critical current of Ref. \cite{boshier13}. On the other hand, the minimum barrier frequency $f^{(2)}_b$
above which there are no more $\phi$ turning points, with the corresponding lack of any DBP regime
(0.2775 Hz for GP results), corresponds in the inset
of Fig.~\ref{figu8} to the transition from black circles to red stars.

The increasing discrepancies between the GP simulation results and those of the SMMB
 above certain asymmetry shown in Fig. \ref{figu6} arise, as already pointed out, from the
use of the parameters of the symmetric system in the SMMB. 
\begin{table}
\caption{ETM Model parameters for different positions of the barriers.
The values are given in nK except
for the particle imbalance $Z_0$.}
\begin{ruledtabular}
\begin{tabular}{l|ccccccccc}
Parameter  & $ \theta=0 $ &   $ \theta=\pi /16 $ &  $ \theta=\pi /8 $  &  $ \theta=\pi /5 $  \\
\hline
$  \tilde{A}$ &  $  0  $ & $  4.734 $    & $  10.06 $  & $ 15.95 $  \\
\hline
$  \tilde{U}_a$  &   $  9.655 \times 10^{-3}  $ & $  9.769 \times 10^{-3}  $    & $  1.000  \times 10^{-2}  $   & $  1.030  \times 10^{-2}  $\\
\hline
$  \tilde{B}$  &   $  0  $ & $  7.590 \times 10^{-4}  $    & $  1.823  \times 10^{-3}  $   & $  4.278  \times 10^{-3}  $\\
\hline
$  J_a$  &   $ 1.495  \times 10^{-2}  $ & $ 1.474  \times 10^{-2}  $    & $  1.395  \times 10^{-2}  $  & $ 1.109  \times 10^{-2}  $ \\
\hline
$  J_b$  &   $ 0 $ & $ 2.512  \times 10^{-3}  $    & $  5.693  \times 10^{-3}  $  & $ 1.226 \times 10^{-2}  $ \\
\hline
$  I$  &   $  7.068 \times 10^{-8}  $ & $  7.256 \times 10^{-8}  $    & $  7.885  \times 10^{-8}  $   & $  9.606  \times 10^{-8}  $\\
\hline
$ Z_0  $  &   $  0  $ & $  0.1615  $    & $  0.323  $   & $  0.516  $\\
\end{tabular}
\end{ruledtabular}
\label{table}
\end{table}
A first attempt to quantify the extent to which such an approximation could affect the SMMB results,
can be evaluated by appreciating the differences in the value of the ETM model parameters in
 Table \ref{table} with respect to those of the symmetric case. 
Note that the only nonvanishing parameters of the symmetric configuration
 are those of the
SMMB, $  \tilde{U}_a=\tilde{U}$ and $J_a=J$, and the neglected parameter $I$.
On the other hand, although all the remaining parameters become finite for asymmetric configurations,
 for $\theta\le\pi/8$ only a little effect of such asymmetries on the SMMB accuracy
should be expected to occur
since,  $  \tilde{U}_a$ and $  J_a$ vary less than 10 \%, we have $Z_0J_b\ll J_a$, and 
$\tilde{A}\simeq\tilde{U}_aNZ_0$ (cf. Eq. (\ref{min})) is well fulfilled in Table \ref{table}.
This explains the good agreement between GP and SMMB results observed in Fig.~\ref{figu8}
and also the agreement with experimental data reported in Ref. \cite{boshier13}.

Now, the simplest improvement to the SMMB in order to take into account the evolving trap
configuration may be provided by an immediate generalization of the ETM model
with its parameters depending on the instantaneous trap asymmetry, i.e., for 
a time-dependent $\theta$.
\begin{figure}
\includegraphics{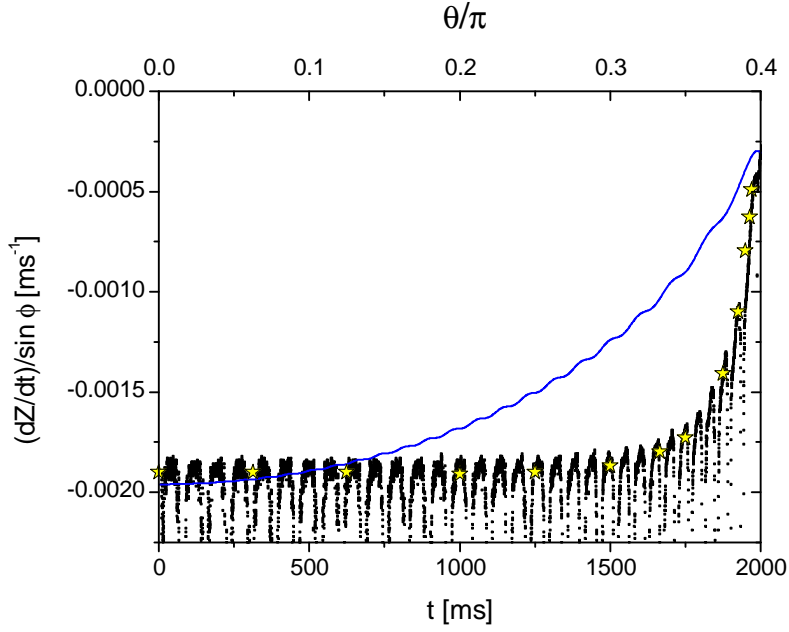}
\caption{(Color online) $(dZ/dt)/\sin\phi$ calculated from GP simulation results (black dots), from
the ETM model,
$-\sqrt{1-Z^2}(J_a + Z J_b)/\hbar    +  (I/\hbar) N  \,  (1 - Z^2) \sin (2 \phi)/\sin \phi$
(yellow stars), and from the SMMB, $-\sqrt{1-Z^2}J/\hbar$ (blue solid line).
The barrier velocity corresponds to $f_b=0.1$ Hz.}
\label{figu9}
\end{figure}
To test such a possibility in the DBP regime, we depict in Figs.~\ref{figu9}
 and \ref{figu10} the derivatives of imbalance and 
phase difference as functions of time 
arising from GP simulations for $f_b=0.1$ Hz, and compare such results with those given by the SMMB,
 and with the corresponding values arising from the ETM equations (\ref{imbe})-(\ref{phaseemo})
for a variable (time-dependent) $\theta$ with $Z$ and $\phi$ taken from the GP simulation results. 
Then we may observe in Fig.~\ref{figu9} that the SMMB prediction for the imbalance derivative
(blue solid line) shows 
an increasing departure from the GP results for $\theta$ above 0.15 $\pi$. 
On the other hand, the results arising
from the ETM model (yellow stars) show a better agreement with the GP simulation values
for the whole time evolution. 
\begin{figure}
\includegraphics{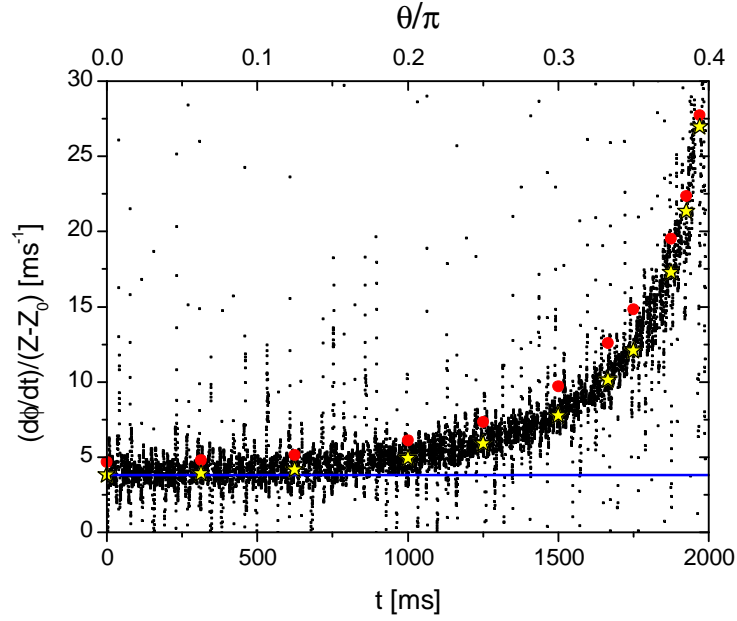}
\caption{(Color online) $(d\phi/dt)/(Z-Z_0)$ calculated from GP simulation results (black dots),
from the ETM model, $N(\tilde{U}_a+Z\tilde{B})/\hbar$ (yellow stars), 
from the plain TM model, $NU_a/\hbar$ (red circles), 
and from the SMMB, $N\tilde{U}/\hbar$ (blue solid line).
The barrier velocity corresponds to $f_b=0.1$ Hz.}
\label{figu10}
\end{figure}

As regards the calculations of the phase derivative, 
first we want to remark that we have found that the
hopping terms (terms in $J_a$, $J_b$, $I$, and  $J$)
in the equations (\ref{phasee}) for TM, (\ref{phaseemo}) for ETM, and (\ref{sphasee}) of
the SMMB turn out to be all negligible with respect to the on-site energy dependent terms, 
so we have disregarded their contribution in the calculations corresponding to
Fig.~\ref{figu10}. Thus, again one finds that the SMMB results
increasingly differ from the GP results for $\theta$ above 0.15 $\pi$, and also that the ETM calculation
clearly improves the agreement with the simulation results. 
In addition, it is possible in this case to test the accuracy of
the ETM model results versus those of the plain TM model, as observed
in Fig.~\ref{figu10}.

\section{Conclusion}\label{sec5}
We have developed a two-mode model of a Bose-Einstein condensate in an asymmetric double-well.
Taking into account effective interaction effects as proposed previously for symmetric configurations, we have introduced corrections to such a model 
in order to improve the agreement with simulation results. Thus, we applied this formalism to a recently
explored experimental setting of a toroidal trap split into an asymmetric double-well condensate
by means of a pair of radial barriers. We have found that the qualitative agreement with the GP simulation results arising from the plain asymmetric TM model, became enhanced to the extent of
an excellent concordance when the effective interaction effects were considered, for practically the whole range of asymmetric configurations. 

We have explored the range of validity of the 
simplest theoretical model for moving barriers, 
previously utilized to describe experimental results,
finding that it should fail for larger departures from the symmetric configuration than
those considered in the experiment, and also for the running-phase regime of higher 
barrier frequencies. Finally, we have performed an analysis of a possible improvement 
of this model, which consists in an immediate generalization of the asymmetric ETM model
to a moving-barrier configuration, simply by employing parameters depending on the
instantaneous trap asymmetry. Our results obtained for the DBP regime pave the way for
further studying more complex dynamics driven by different kinds of barrier movements.

\appendix*
\section{TM model parameters}

The TM model parameters read,

\begin{equation}
\varepsilon_{1} = \int d^3r\,\,  \psi_{1}({\bf r}) \left[
-\frac{ \hbar^2 }{2 m}{\bf \nabla}^2 +
V_{\rm{trap}}({\bf r})\right]  \psi_{1}({\bf r})
\label{epsR}
\end{equation}

\begin{equation}
\varepsilon_{2} = \int d^3r\,\,  \psi_{2}({\bf r}) \left[
-\frac{ \hbar^2 }{2 m}{\bf \nabla}^2 +
V_{\rm{trap}}({\bf r})\right]  \psi_{2}({\bf r})
\label{epsL}
\end{equation}

\begin{equation}
K= -\int d^3r\,\, \psi_{1}({\bf r}) \left[
-\frac{ \hbar^2 }{2 m}{\bf \nabla}^2  +
V_{\rm{trap}}({\bf r})\right]  \psi_{2}({\bf r})
\label{jota0}
\end{equation}

\begin{equation}
U_{1}= g \int d^3r\,\,  \psi_{1}^4({\bf r})
\label{U0R}
\end{equation}

\begin{equation}
U_{2}= g \int d^3r\,\,  \psi_{2}^4({\bf r})
\label{U0L}
\end{equation}

\begin{equation}
F_{12}=   -g\int d^3r\,\,  \psi_{1}^3({\bf r})
 \psi_{2}({\bf r})
\label{FRL}
\end{equation}

\begin{equation}
F_{21}=   -g\int d^3r\,\,  \psi_{1}({\bf r})
 \psi_{2}^3({\bf r})
\label{FLR}
\end{equation}

\begin{equation}
I= g  \int d^3r\,\,   \psi_{1}^2({\bf r}) \,  \psi_{2}^2({\bf r}) .
\label{ijota}
\end{equation}
We note that according to the definitions (\ref{A}) and (\ref{Ua}), $A$ and $U_a$ turn out to be
on-site energy dependent parameters, as they are built from the on-site energy parameters
 $\varepsilon_k$ and $U_k$. On the other hand, $K$ is the standard hopping coefficient,
while the remaining hopping parameters
$ F_{12} $, $ F_{21} $ and $I$  have been first  introduced by Ananikian and Bergeman 
  in Ref.~\cite{anan06}. Particularly, $ N F_{12} $ and $ N F_{21} $  can 
be interpreted 
 as additional contributions to the  tunnelling obtained from a modified  Hamiltonian that includes
  the interaction term as   an  effective potential \cite{jia08}.  We
remark that in both references the authors assume symmetric traps with $  F_{12} =  F_{21} $,
a parameter
which in our case would only contribute to the hopping parameter $J_a$ (\ref{Ja}). 
On the other hand, in an asymmetric trap,
the difference between $F_{21} $ and $  F_{12}$
gives rise to the additional hopping parameter $J_b$ (\ref{Jb}).
As regards the parameter $I$, it corresponds to atom-pair tunnelling processes and, although 
it turns out to be in most cases negligible, it was recently shown that for enough strong interaction
regimes a quantum phase transition driven by such atom-pair tunnelling events 
should be expected to take place \cite{Liu}.
\acknowledgments
 HMC and DMJ  acknowledge CONICET for financial support under Grant Nos. 
 PIP 11420100100083 and PIP 11420090100243 , respectively.
\providecommand{\noopsort}[1]{}\providecommand{\singleletter}[1]{#1}%
%


\begin{thebibliography}{22}%
\makeatletter
\providecommand \@ifxundefined [1]{%
 \@ifx{#1\undefined}
}%
\providecommand \@ifnum [1]{%
 \ifnum #1\expandafter \@firstoftwo
 \else \expandafter \@secondoftwo
 \fi
}%
\providecommand \@ifx [1]{%
 \ifx #1\expandafter \@firstoftwo
 \else \expandafter \@secondoftwo
 \fi
}%
\providecommand \natexlab [1]{#1}%
\providecommand \enquote  [1]{``#1''}%
\providecommand \bibnamefont  [1]{#1}%
\providecommand \bibfnamefont [1]{#1}%
\providecommand \citenamefont [1]{#1}%
\providecommand \href@noop [0]{\@secondoftwo}%
\providecommand \href [0]{\begingroup \@sanitize@url \@href}%
\providecommand \@href[1]{\@@startlink{#1}\@@href}%
\providecommand \@@href[1]{\endgroup#1\@@endlink}%
\providecommand \@sanitize@url [0]{\catcode `\\12\catcode `\$12\catcode
  `\&12\catcode `\#12\catcode `\^12\catcode `\_12\catcode `\%12\relax}%
\providecommand \@@startlink[1]{}%
\providecommand \@@endlink[0]{}%
\providecommand \url  [0]{\begingroup\@sanitize@url \@url }%
\providecommand \@url [1]{\endgroup\@href {#1}{\urlprefix }}%
\providecommand \urlprefix  [0]{URL }%
\providecommand \Eprint [0]{\href }%
\providecommand \doibase [0]{http://dx.doi.org/}%
\providecommand \selectlanguage [0]{\@gobble}%
\providecommand \bibinfo  [0]{\@secondoftwo}%
\providecommand \bibfield  [0]{\@secondoftwo}%
\providecommand \translation [1]{[#1]}%
\providecommand \BibitemOpen [0]{}%
\providecommand \bibitemStop [0]{}%
\providecommand \bibitemNoStop [0]{.\EOS\space}%
\providecommand \EOS [0]{\spacefactor3000\relax}%
\providecommand \BibitemShut  [1]{\csname bibitem#1\endcsname}%
\let\auto@bib@innerbib\@empty
\bibitem [{\citenamefont {Smerzi}\ \emph {et~al.}(1997)\citenamefont {Smerzi},
  \citenamefont {Fantoni}, \citenamefont {Giovanazzi},\ and\ \citenamefont
  {Shenoy}}]{smerzi97}%
  \BibitemOpen
  \bibfield  {author} {\bibinfo {author} {\bibfnamefont {A.}~\bibnamefont
  {Smerzi}}, \bibinfo {author} {\bibfnamefont {S.}~\bibnamefont {Fantoni}},
  \bibinfo {author} {\bibfnamefont {S.}~\bibnamefont {Giovanazzi}}, \ and\
  \bibinfo {author} {\bibfnamefont {S.~R.}\ \bibnamefont {Shenoy}},\
  }\href@noop {} {\bibfield  {journal} {\bibinfo  {journal} {Phys. Rev. Lett}\
  }\textbf {\bibinfo {volume} {79}},\ \bibinfo {pages} {4950} (\bibinfo {year}
  {1997})}\BibitemShut {NoStop}%
\bibitem [{\citenamefont {Raghavan}\ \emph {et~al.}(1999)\citenamefont
  {Raghavan}, \citenamefont {Smerzi}, \citenamefont {Fantoni},\ and\
  \citenamefont {Shenoy}}]{ragh99}%
  \BibitemOpen
  \bibfield  {author} {\bibinfo {author} {\bibfnamefont {S.}~\bibnamefont
  {Raghavan}}, \bibinfo {author} {\bibfnamefont {A.}~\bibnamefont {Smerzi}},
  \bibinfo {author} {\bibfnamefont {S.}~\bibnamefont {Fantoni}}, \ and\
  \bibinfo {author} {\bibfnamefont {S.~R.}\ \bibnamefont {Shenoy}},\
  }\href@noop {} {\bibfield  {journal} {\bibinfo  {journal} {Phys. Rev. A}\
  }\textbf {\bibinfo {volume} {59}},\ \bibinfo {pages} {620} (\bibinfo {year}
  {1999})}\BibitemShut {NoStop}%
\bibitem [{\citenamefont {Ananikian}\ and\ \citenamefont
  {Bergeman}(2006)}]{anan06}%
  \BibitemOpen
  \bibfield  {author} {\bibinfo {author} {\bibfnamefont {D.}~\bibnamefont
  {Ananikian}}\ and\ \bibinfo {author} {\bibfnamefont {T.}~\bibnamefont
  {Bergeman}},\ }\href@noop {} {\bibfield  {journal} {\bibinfo  {journal}
  {Phys. Rev. A}\ }\textbf {\bibinfo {volume} {73}},\ \bibinfo {pages} {013604}
  (\bibinfo {year} {2006})}\BibitemShut {NoStop}%
\bibitem [{\citenamefont {{\relax Xin Yan Jia}}\ \emph
  {et~al.}(2008)\citenamefont {{\relax Xin Yan Jia}}, \citenamefont {{\relax
  WeiDong Li}},\ and\ \citenamefont {Liang}}]{jia08}%
  \BibitemOpen
  \bibfield  {author} {\bibinfo {author} {\bibnamefont {{\relax Xin Yan Jia}}},
  \bibinfo {author} {\bibnamefont {{\relax WeiDong Li}}}, \ and\ \bibinfo
  {author} {\bibfnamefont {J.~Q.}\ \bibnamefont {Liang}},\ }\href@noop {}
  {\bibfield  {journal} {\bibinfo  {journal} {Phys. Rev. A}\ }\textbf {\bibinfo
  {volume} {78}},\ \bibinfo {pages} {023613} (\bibinfo {year}
  {2008})}\BibitemShut {NoStop}%
\bibitem [{\citenamefont {Albiez}\ \emph {et~al.}(2005)\citenamefont {Albiez},
  \citenamefont {Gati}, \citenamefont {F{\"{o}}lling}, \citenamefont
  {Hunsmann}, \citenamefont {Cristiani},\ and\ \citenamefont
  {Oberthaler}}]{albiez05}%
  \BibitemOpen
  \bibfield  {author} {\bibinfo {author} {\bibfnamefont {M.}~\bibnamefont
  {Albiez}}, \bibinfo {author} {\bibfnamefont {R.}~\bibnamefont {Gati}},
  \bibinfo {author} {\bibfnamefont {J.}~\bibnamefont {F{\"{o}}lling}}, \bibinfo
  {author} {\bibfnamefont {S.}~\bibnamefont {Hunsmann}}, \bibinfo {author}
  {\bibfnamefont {M.}~\bibnamefont {Cristiani}}, \ and\ \bibinfo {author}
  {\bibfnamefont {M.~K.}\ \bibnamefont {Oberthaler}},\ }\href@noop {}
  {\bibfield  {journal} {\bibinfo  {journal} {Phys. Rev. Lett.}\ }\textbf
  {\bibinfo {volume} {95}},\ \bibinfo {pages} {010402} (\bibinfo {year}
  {2005})}\BibitemShut {NoStop}%
\bibitem [{\citenamefont {Mel{\'{e}}-Messeguer}\ \emph
  {et~al.}(2011)\citenamefont {Mel{\'{e}}-Messeguer}, \citenamefont
  {Juli\'a-D\'{\i}az}, \citenamefont {Guilleumas}, \citenamefont {Polls},\ and\
  \citenamefont {Sanpera}}]{mele11}%
  \BibitemOpen
  \bibfield  {author} {\bibinfo {author} {\bibfnamefont {M.}~\bibnamefont
  {Mel{\'{e}}-Messeguer}}, \bibinfo {author} {\bibfnamefont {B.}~\bibnamefont
  {Juli\'a-D\'{\i}az}}, \bibinfo {author} {\bibfnamefont {M.}~\bibnamefont
  {Guilleumas}}, \bibinfo {author} {\bibfnamefont {A.}~\bibnamefont {Polls}}, \
  and\ \bibinfo {author} {\bibfnamefont {A.}~\bibnamefont {Sanpera}},\
  }\href@noop {} {\bibfield  {journal} {\bibinfo  {journal} {New J. Phys.}\
  }\textbf {\bibinfo {volume} {13}},\ \bibinfo {pages} {033012} (\bibinfo
  {year} {2011})}\BibitemShut {NoStop}%
\bibitem [{\citenamefont {Abad}\ \emph
  {et~al.}(2011{\natexlab{a}})\citenamefont {Abad}, \citenamefont {Guilleumas},
  \citenamefont {Mayol}, \citenamefont {Pi},\ and\ \citenamefont
  {Jezek}}]{abad11}%
  \BibitemOpen
  \bibfield  {author} {\bibinfo {author} {\bibfnamefont {M.}~\bibnamefont
  {Abad}}, \bibinfo {author} {\bibfnamefont {M.}~\bibnamefont {Guilleumas}},
  \bibinfo {author} {\bibfnamefont {R.}~\bibnamefont {Mayol}}, \bibinfo
  {author} {\bibfnamefont {M.}~\bibnamefont {Pi}}, \ and\ \bibinfo {author}
  {\bibfnamefont {D.~M.}\ \bibnamefont {Jezek}},\ }\href@noop {} {\bibfield
  {journal} {\bibinfo  {journal} {Europhys. Lett.}\ }\textbf {\bibinfo {volume}
  {94}},\ \bibinfo {pages} {10004} (\bibinfo {year}
  {2011}{\natexlab{a}})}\BibitemShut {NoStop}%
\bibitem [{\citenamefont {Mayteevarunyoo}\ \emph {et~al.}(2008)\citenamefont
  {Mayteevarunyoo}, \citenamefont {Malomed},\ and\ \citenamefont
  {Dong}}]{doublewell}%
  \BibitemOpen
  \bibfield  {author} {\bibinfo {author} {\bibfnamefont {T.}~\bibnamefont
  {Mayteevarunyoo}}, \bibinfo {author} {\bibfnamefont {B.~A.}\ \bibnamefont
  {Malomed}}, \ and\ \bibinfo {author} {\bibfnamefont {G.}~\bibnamefont
  {Dong}},\ }\href@noop {} {\bibfield  {journal} {\bibinfo  {journal} {Phys.
  Rev. A}\ }\textbf {\bibinfo {volume} {78}},\ \bibinfo {pages} {053601}
  (\bibinfo {year} {2008})}\BibitemShut {NoStop}%
\bibitem [{\citenamefont {Xiong}\ \emph {et~al.}(2009)\citenamefont {Xiong},
  \citenamefont {Gong}, \citenamefont {Pu}, \citenamefont {Bao},\ and\
  \citenamefont {Li}}]{xiong}%
  \BibitemOpen
  \bibfield  {author} {\bibinfo {author} {\bibfnamefont {B.}~\bibnamefont
  {Xiong}}, \bibinfo {author} {\bibfnamefont {J.}~\bibnamefont {Gong}},
  \bibinfo {author} {\bibfnamefont {H.}~\bibnamefont {Pu}}, \bibinfo {author}
  {\bibfnamefont {W.}~\bibnamefont {Bao}}, \ and\ \bibinfo {author}
  {\bibfnamefont {B.}~\bibnamefont {Li}},\ }\href@noop {} {\bibfield  {journal}
  {\bibinfo  {journal} {Phys. Rev. A}\ }\textbf {\bibinfo {volume} {79}},\
  \bibinfo {pages} {013626} (\bibinfo {year} {2009})}\BibitemShut {NoStop}%
\bibitem [{\citenamefont {{\relax Qi Zhou}}\ \emph {et~al.}(2011)\citenamefont
  {{\relax Qi Zhou}}, \citenamefont {Porto},\ and\ \citenamefont {{\relax S.
  Das Sarma}}}]{zhou}%
  \BibitemOpen
  \bibfield  {author} {\bibinfo {author} {\bibnamefont {{\relax Qi Zhou}}},
  \bibinfo {author} {\bibfnamefont {J.~V.}\ \bibnamefont {Porto}}, \ and\
  \bibinfo {author} {\bibnamefont {{\relax S. Das Sarma}}},\ }\href@noop {}
  {\bibfield  {journal} {\bibinfo  {journal} {Phys. Rev. A}\ }\textbf {\bibinfo
  {volume} {84}},\ \bibinfo {pages} {031607} (\bibinfo {year}
  {2011})}\BibitemShut {NoStop}%
\bibitem [{\citenamefont {Cui}\ \emph {et~al.}(2010)\citenamefont {Cui},
  \citenamefont {Wang},\ and\ \citenamefont {Yi}}]{cui}%
  \BibitemOpen
  \bibfield  {author} {\bibinfo {author} {\bibfnamefont {B.}~\bibnamefont
  {Cui}}, \bibinfo {author} {\bibfnamefont {L.~C.}\ \bibnamefont {Wang}}, \
  and\ \bibinfo {author} {\bibfnamefont {X.~X.}\ \bibnamefont {Yi}},\
  }\href@noop {} {\bibfield  {journal} {\bibinfo  {journal} {Phys. Rev. A}\
  }\textbf {\bibinfo {volume} {82}},\ \bibinfo {pages} {062105} (\bibinfo
  {year} {2010})}\BibitemShut {NoStop}%
\bibitem [{\citenamefont {Abad}\ \emph
  {et~al.}(2011{\natexlab{b}})\citenamefont {Abad}, \citenamefont {Guilleumas},
  \citenamefont {Mayol}, \citenamefont {Pi},\ and\ \citenamefont
  {Jezek}}]{gui}%
  \BibitemOpen
  \bibfield  {author} {\bibinfo {author} {\bibfnamefont {M.}~\bibnamefont
  {Abad}}, \bibinfo {author} {\bibfnamefont {M.}~\bibnamefont {Guilleumas}},
  \bibinfo {author} {\bibfnamefont {R.}~\bibnamefont {Mayol}}, \bibinfo
  {author} {\bibfnamefont {M.}~\bibnamefont {Pi}}, \ and\ \bibinfo {author}
  {\bibfnamefont {D.~M.}\ \bibnamefont {Jezek}},\ }\href@noop {} {\bibfield
  {journal} {\bibinfo  {journal} {Phys. Rev. A}\ }\textbf {\bibinfo {volume}
  {84}},\ \bibinfo {pages} {035601} (\bibinfo {year}
  {2011}{\natexlab{b}})}\BibitemShut {NoStop}%
\bibitem [{\citenamefont {Jezek}\ and\ \citenamefont {Cataldo}(2013)}]{je13}%
  \BibitemOpen
  \bibfield  {author} {\bibinfo {author} {\bibfnamefont {D.~M.}\ \bibnamefont
  {Jezek}}\ and\ \bibinfo {author} {\bibfnamefont {H.~M.}\ \bibnamefont
  {Cataldo}},\ }\href@noop {} {\bibfield  {journal} {\bibinfo  {journal} {Phys.
  Rev. A}\ }\textbf {\bibinfo {volume} {88}},\ \bibinfo {pages} {013636}
  (\bibinfo {year} {2013})}\BibitemShut {NoStop}%
\bibitem [{\citenamefont {Ryu}\ \emph {et~al.}(2013)\citenamefont {Ryu},
  \citenamefont {Blackburn}, \citenamefont {Blinova},\ and\ \citenamefont
  {Boshier}}]{boshier13}%
  \BibitemOpen
  \bibfield  {author} {\bibinfo {author} {\bibfnamefont {C.}~\bibnamefont
  {Ryu}}, \bibinfo {author} {\bibfnamefont {P.~W.}\ \bibnamefont {Blackburn}},
  \bibinfo {author} {\bibfnamefont {A.~A.}\ \bibnamefont {Blinova}}, \ and\
  \bibinfo {author} {\bibfnamefont {M.~G.}\ \bibnamefont {Boshier}},\
  }\href@noop {} {\bibfield  {journal} {\bibinfo  {journal} {Phys. Rev. Lett.}\
  }\textbf {\bibinfo {volume} {111}},\ \bibinfo {pages} {205301} (\bibinfo
  {year} {2013})}\BibitemShut {NoStop}%
\bibitem [{\citenamefont {Sato}(2013)}]{sato}%
  \BibitemOpen
  \bibfield  {author} {\bibinfo {author} {\bibfnamefont {Y.}~\bibnamefont
  {Sato}},\ }\href@noop {} {\bibfield  {journal} {\bibinfo  {journal}
  {Physics}\ }\textbf {\bibinfo {volume} {6}},\ \bibinfo {pages} {123}
  (\bibinfo {year} {2013})}\BibitemShut {NoStop}%
\bibitem [{\citenamefont {Jendrzejewski}\ \emph {et~al.}(2014)\citenamefont
  {Jendrzejewski}, \citenamefont {Eckel}, \citenamefont {Murray}, \citenamefont
  {Lanier}, \citenamefont {Edwards}, \citenamefont {Lobb},\ and\ \citenamefont
  {Campbell}}]{jen14}%
  \BibitemOpen
  \bibfield  {author} {\bibinfo {author} {\bibfnamefont {F.}~\bibnamefont
  {Jendrzejewski}}, \bibinfo {author} {\bibfnamefont {S.}~\bibnamefont
  {Eckel}}, \bibinfo {author} {\bibfnamefont {N.}~\bibnamefont {Murray}},
  \bibinfo {author} {\bibfnamefont {C.}~\bibnamefont {Lanier}}, \bibinfo
  {author} {\bibfnamefont {M.}~\bibnamefont {Edwards}}, \bibinfo {author}
  {\bibfnamefont {C.~J.}\ \bibnamefont {Lobb}}, \ and\ \bibinfo {author}
  {\bibfnamefont {G.~K.}\ \bibnamefont {Campbell}},\ }\href@noop {} {\bibfield
  {journal} {\bibinfo  {journal} {Phys. Rev. Lett.}\ }\textbf {\bibinfo
  {volume} {113}},\ \bibinfo {pages} {045305} (\bibinfo {year}
  {2014})}\BibitemShut {NoStop}%
\bibitem [{\citenamefont {Jezek}\ \emph {et~al.}(2013)\citenamefont {Jezek},
  \citenamefont {Capuzzi},\ and\ \citenamefont {Cataldo}}]{cap13}%
  \BibitemOpen
  \bibfield  {author} {\bibinfo {author} {\bibfnamefont {D.~M.}\ \bibnamefont
  {Jezek}}, \bibinfo {author} {\bibfnamefont {P.}~\bibnamefont {Capuzzi}}, \
  and\ \bibinfo {author} {\bibfnamefont {H.~M.}\ \bibnamefont {Cataldo}},\
  }\href@noop {} {\bibfield  {journal} {\bibinfo  {journal} {Phys. Rev. A}\
  }\textbf {\bibinfo {volume} {87}},\ \bibinfo {pages} {053625} (\bibinfo
  {year} {2013})}\BibitemShut {NoStop}%
\bibitem [{\citenamefont {LeBlanc}\ \emph {et~al.}(2011)\citenamefont
  {LeBlanc}, \citenamefont {Bardon}, \citenamefont {McKeever}, \citenamefont
  {Extavour}, \citenamefont {Jervis}, \citenamefont {Thywissen}, \citenamefont
  {Piazza},\ and\ \citenamefont {Smerzi}}]{LeBlanc11}%
  \BibitemOpen
  \bibfield  {author} {\bibinfo {author} {\bibfnamefont {L.~J.}\ \bibnamefont
  {LeBlanc}}, \bibinfo {author} {\bibfnamefont {A.~B.}\ \bibnamefont {Bardon}},
  \bibinfo {author} {\bibfnamefont {J.}~\bibnamefont {McKeever}}, \bibinfo
  {author} {\bibfnamefont {M.~H.~T.}\ \bibnamefont {Extavour}}, \bibinfo
  {author} {\bibfnamefont {D.}~\bibnamefont {Jervis}}, \bibinfo {author}
  {\bibfnamefont {J.~H.}\ \bibnamefont {Thywissen}}, \bibinfo {author}
  {\bibfnamefont {F.}~\bibnamefont {Piazza}}, \ and\ \bibinfo {author}
  {\bibfnamefont {A.}~\bibnamefont {Smerzi}},\ }\href@noop {} {\bibfield
  {journal} {\bibinfo  {journal} {Phys. Rev. Lett}\ }\textbf {\bibinfo {volume}
  {106}},\ \bibinfo {pages} {025302} (\bibinfo {year} {2011})}\BibitemShut
  {NoStop}%
\bibitem [{\citenamefont {Wright}\ \emph {et~al.}(2000)\citenamefont {Wright},
  \citenamefont {Arlt},\ and\ \citenamefont {Dholakia}}]{lag}%
  \BibitemOpen
  \bibfield  {author} {\bibinfo {author} {\bibfnamefont {E.~M.}\ \bibnamefont
  {Wright}}, \bibinfo {author} {\bibfnamefont {J.}~\bibnamefont {Arlt}}, \ and\
  \bibinfo {author} {\bibfnamefont {K.}~\bibnamefont {Dholakia}},\ }\href@noop
  {} {\bibfield  {journal} {\bibinfo  {journal} {Phys. Rev. A}\ }\textbf
  {\bibinfo {volume} {63}},\ \bibinfo {pages} {013608} (\bibinfo {year}
  {2000})}\BibitemShut {NoStop}%
\bibitem [{\citenamefont {Castin}\ and\ \citenamefont {Dum}(1999)}]{castin}%
  \BibitemOpen
  \bibfield  {author} {\bibinfo {author} {\bibfnamefont {Y.}~\bibnamefont
  {Castin}}\ and\ \bibinfo {author} {\bibfnamefont {R.}~\bibnamefont {Dum}},\
  }\href@noop {} {\bibfield  {journal} {\bibinfo  {journal} {Eur. Phys. J. D}\
  }\textbf {\bibinfo {volume} {7}},\ \bibinfo {pages} {399} (\bibinfo {year}
  {1999})}\BibitemShut {NoStop}%
\bibitem [{\citenamefont {Giovanazzi}\ \emph {et~al.}(2000)\citenamefont
  {Giovanazzi}, \citenamefont {Smerzi},\ and\ \citenamefont
  {Fantoni}}]{giova00}%
  \BibitemOpen
  \bibfield  {author} {\bibinfo {author} {\bibfnamefont {S.}~\bibnamefont
  {Giovanazzi}}, \bibinfo {author} {\bibfnamefont {A.}~\bibnamefont {Smerzi}},
  \ and\ \bibinfo {author} {\bibfnamefont {S.}~\bibnamefont {Fantoni}},\
  }\href@noop {} {\bibfield  {journal} {\bibinfo  {journal} {Phys. Rev. Lett}\
  }\textbf {\bibinfo {volume} {84}},\ \bibinfo {pages} {4521} (\bibinfo {year}
  {2000})}\BibitemShut {NoStop}%
\bibitem [{\citenamefont {{\relax J-L Liu}}\ and\ \citenamefont {{\relax J-Q
  Liang}}(2011)}]{Liu}%
  \BibitemOpen
  \bibfield  {author} {\bibinfo {author} {\bibnamefont {{\relax J-L Liu}}}\
  and\ \bibinfo {author} {\bibnamefont {{\relax J-Q Liang}}},\ }\href@noop {}
  {\bibfield  {journal} {\bibinfo  {journal} {J. Phys. B: At. Mol. Opt. Phys.}\
  }\textbf {\bibinfo {volume} {44}},\ \bibinfo {pages} {025101} (\bibinfo
  {year} {2011})}\BibitemShut {NoStop}%
\end{thebibliography}
\end{document}